\documentclass[aps,amsmath,amssymb,preprintnumbers,superscriptaddress,twocolumn,floatfix,superscriptaddress,nofootinbib]{revtex4-2}
\usepackage{amsmath}
\usepackage{amsfonts}
\usepackage{amssymb}
\usepackage{tikz}
\usepackage{graphicx}
\usepackage{microtype}
\definecolor{cLink}{RGB}{1, 1, 128}
\definecolor{cURL}{RGB}{1, 1, 128}
\definecolor{cCite}{RGB}{1, 1, 128}
\usepackage[colorlinks=true, linkcolor=cLink, urlcolor=cURL, citecolor=cCite]{hyperref}
\usepackage{booktabs}
\usepackage{mathtools}
\allowdisplaybreaks

\newcommand{\MPleleven}{M_{\textrm{Pl}}^{\textrm{11D}}}
\newcommand{\MPleight}{M_{\textrm{Pl}}^{\textrm{8D}}}
\newcommand{\MPlfive}{M_{\textrm{Pl}}^{\textrm{5D}}}
\newcommand{\MPlfour}{M_{\textrm{Pl}}^{\textrm{4D}}}
\newcommand{\vol}[2][]{\mathcal{V}^{#1}_{#2}}
\newcommand{\Veleven}{\mathcal{V}_{X}^{\textrm{11D}}}
\newcommand{\Vfour}{\mathcal{V}_{X}^{\IIA}}
\newcommand{\IIA}{\textrm{IIA}}
\newcommand{\IIB}{\textrm{IIB}}
\newcommand{\QC}{\textrm{QC}}
\newcommand{\SYZ}{\textrm{SYZ}}
\makeatletter
\newcommand\footnoteref[1]{\protected@xdef\@thefnmark{\ref{#1}}\@footnotemark}
\makeatother
\definecolor{pureblue}{HTML}{000080}
\definecolor{purered}{HTML}{ff0000}
\begin{document}

\preprint{ZMP-HH/21-25}

\title{Membrane Limits in Quantum Gravity}
	
\author{Rafael \'Alvarez-Garc\'ia}
\affiliation{%
 II. Institut f\"ur Theoretische Physik, Universit\"at Hamburg\\
 Luruper Chaussee 149, 22607 Hamburg, Germany
 }
\author{Daniel Kl\"awer}
\affiliation{%
 II. Institut f\"ur Theoretische Physik, Universit\"at Hamburg\\
 Luruper Chaussee 149, 22607 Hamburg, Germany
 }
\author{Timo Weigand}
\affiliation{%
 II. Institut f\"ur Theoretische Physik, Universit\"at Hamburg\\
 Luruper Chaussee 149, 22607 Hamburg, Germany
 }
\affiliation{%
 Zentrum f\"ur Mathematische Physik, Universit\"at Hamburg\\
 Bundesstra\ss e 55, 20146 Hamburg, Germany
 }

\begin{abstract}
It is expected that infinite distance limits in the moduli space of quantum gravity are accompanied by a tower of light states. In view of the emergent string conjecture, this tower must either induce a decompactification or correspond to the emergence of a tensionless critical string. We study the consistency conditions implied by this conjecture on the asymptotic behavior of quantum gravity under dimensional reduction. If the emergent string descends from a (2+1)-dimensional membrane in a higher-dimensional theory, we find that such a membrane must parametrically decouple from the Kaluza-Klein scale. We verify this censorship against emergent membrane limits, where the membrane would sit at the Kaluza-Klein scale, in the hypermultiplet moduli space of Calabi-Yau 3-fold compactifications of string/M-theory. At the classical level, a putative membrane limit arises, up to duality, from an M5-brane wrapping the asymptotically shrinking special Lagrangian 3-cycle corresponding to the Strominger-Yau-Zaslow fiber of the Calabi-Yau. We show how quantum corrections in the moduli space obstruct such a limit and instead lead to a decompactification to 11 dimensions, where the role of the M5- and M2-branes are interchanged.
\end{abstract}

\maketitle

\section{\label{sec:intro}Introduction}

Consistency conditions imposed by quantum gravity restrict the space of acceptable effective field theories (EFTs). From the set of EFTs that naively we would have deemed consistent only a small subset can actually be UV completed to a theory of quantum gravity, and therefore populate the \textit{landscape}. The remaining theories run into consistency problems when gravity is taken into account, and are said to be in the \textit{swampland}.

The \textit{Swampland Program} \cite{Vafa:2005ui} aims to delineate the boundary between the \textit{landscape} and the \textit{swampland}, an endeavour which has received lots of attention in recent years. In this context, an ever-growing web of general interconnected conjectures on the nature of quantum gravity has been taking shape; see \cite{Brennan:2017rbf,Palti:2019pca,vanBeest:2021lhn,Grana:2021zvf} for recent reviews. Among these \textit{swampland conjectures} the Swampland Distance Conjecture (SDC) asserts that along infinite distance limits in moduli space a tower of states has to become asymptotically massless, therefore breaking the EFT description. This has been extensively tested in string theory \cite{Blumenhagen:2018nts,Grimm:2018ohb,Grimm:2018cpv,Corvilain:2018lgw,Font:2019cxq,Joshi:2019nzi,Gendler:2020dfp,Grimm:2020cda,Palti:2021ubp,Grimm:2021ikg,Klawer:2021ltm,Alvarez-Garcia:2021mzv}, where asymptotic regions in moduli space exhibit a simplified structure that allows precise checks at a quantitative level. The conjecture has also been motivated by its relation to the Weak Gravity Conjecture~\cite{Klaewer:2016kiy,Palti:2017elp}.

The Emergent String Conjecture \cite{Lee:2019wij} is a refinement of the SDC proposing that infinite distance limits in moduli space either are pure decompactification limits or signal a transition to a duality frame determined by a unique emergent critical weakly coupled string such that $T_{\textrm{str}} \sim M^{2}_{\textrm{KK}}$. The Emergent String Conjecture has been scrutinized in the context of the K\"ahler moduli space of F/M/IIA-theory in 6D/5D/4D in \cite{Lee:2018urn,Lee:2019xtm,Lee:2019wij}, in the complex structure moduli space of F-theory in 8D in \cite{Kulikov,HetFpaper}, and for the 4D $\mathcal{N}=2$ hypermultiplet moduli space in \cite{Marchesano:2019ifh,Baume:2019sry}. The case of M-theory on $G_{2}$ manifolds was treated in \cite{Xu:2020nlh}, while 4D $\mathcal{N}=1$ F-theory was studied in \cite{Lee:2019tst}, with the leading quantum corrections included in \cite{Klaewer:2020lfg}. Their effect, both in the vector multiplet and hypermultiplet moduli spaces, is to remove pathological string limits in which $T_{\textrm{str}} \prec M^{2}_{\textrm{KK}}$.\footnote{By writing $A\precsim B$, we mean that $\lim (A/B)<\infty$, i.e.\ the two quantities either scale in the same way with respect to a limiting parameter or B is parametrically dominating A; similarly, $A\prec B$ means $\lim (A/B) =0$.} Such limits are expected to be impossible in the landscape, as otherwise one could decouple the Kaluza-Klein (KK) scale and obtain genuinely lower-dimensional critical string theories. Complementary viewpoints on the relevance of strings for determining the asymptotics of moduli space have appeared in \cite{Lanza:2020qmt,Lanza:2021udy,Heidenreich:2021yda}.   

In the present work we analyse infinite distance limits in the hypermultiplet moduli space of M-theory on a Calabi-Yau threefold $X$ in which, classically, a distinguished membrane becomes light at the same rate as the KK scale. Such putative limits will be called (emergent) membrane limits. We find that quantum corrections forbid these limits, deflecting them into trajectories in which the membrane always sits at a scale parametrically higher than the KK scale, as is characteristic for a decompactification of the theory.

Our motivation for this work is twofold.
The first motivation is to challenge the Emergent String Conjecture by trying to construct limits in which an object different from a critical string becomes equally light as the KK scale. 
The natural candidate for such an object would indeed be a membrane which is as close to being ``critical'' as possible, in the sense that at least under dimensional reduction it becomes a critical string.
What we have in mind is a situation similar to the one for the M2-brane in eleven-dimensional M-theory. Even if realisable, emergent membrane limits would presumably still qualify as decompactification limits since, unlike a critical string, the membrane is not expected to give rise to a dense tower of particle excitations. However, if they existed, we could start in the interior of the moduli space, where the KK scale and the tension of the M2-brane are comparable, and move to an infinite distance point, where again a (dual) membrane sits at the same scale as the effective KK scale. In this sense, we would encounter essentially the same theory at infinite distance---analogous to what happens for emergent string limits in the hypermultiplet moduli space of Type II string theory \cite{Baume:2019sry}.
We find it intriguing that gravity censors the appearance of such emergent membrane limits in the quantum moduli space.

The second motivation is to investigate 
the implications of the Emergent String Conjecture upon dimensional reduction.
As it turns out, the observed quantum obstruction against a membrane limit is a consequence of the Emergent String Conjecture in the theory obtained by dimensional reduction. More precisely, emergent membrane limits would lead, under further compactification on an M-theory circle, to pathological string limits in which $T_{\textrm{str}} \prec M^{2}_{\textrm{KK}}$. The quantum obstruction to these limits must therefore already be at work for the membranes in the five-dimensional M-theory setting. In this sense, the Emergent String Conjecture acts as a censor against emergent membrane limits in one dimension higher.

\begin{figure}[t!]
    \begin{tikzpicture}
        \node[inner sep=0pt] at (0,0) {\includegraphics[width=0.45\textwidth]{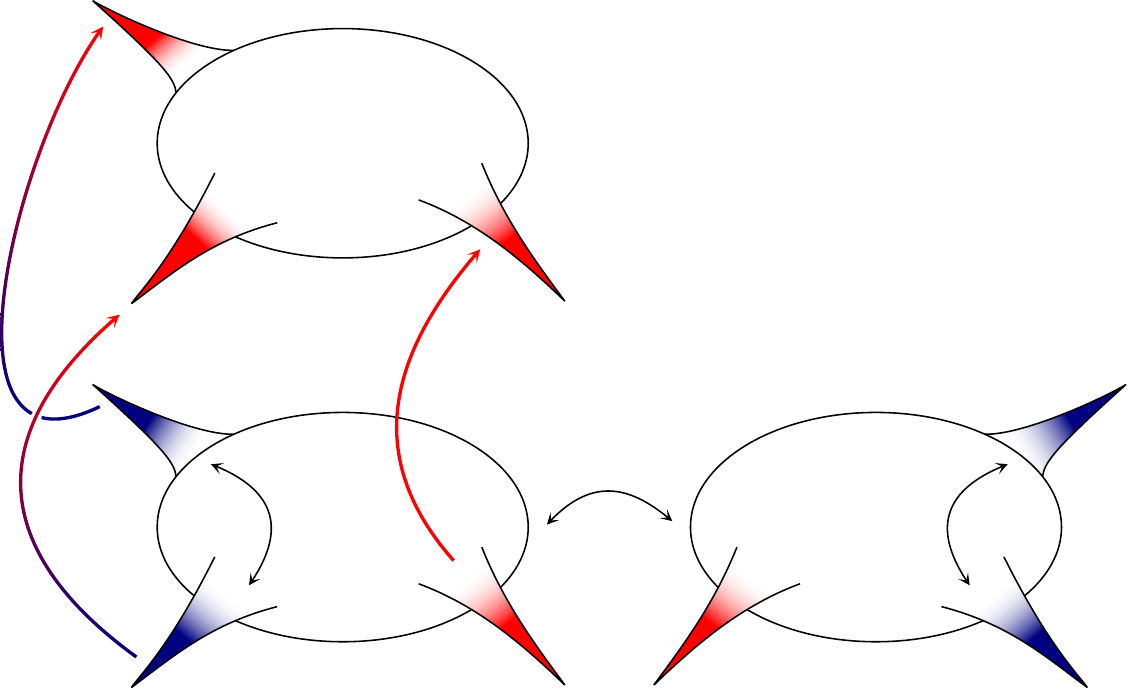}};
        \node at (-0.6,2.5) {$\mathcal{M}_{\textrm{HM}}^{\textrm{M-th}}$};
        \node at (-0.6,-.3) {$\mathcal{M}_{\textrm{HM}}^{\IIA}$};
        \node at (1.2,-.3) {$\mathcal{M}_{\textrm{HM}}^{\IIB}$};
        \draw[pureblue,fill=pureblue] (1,2.1) circle (3pt);
        \node[align=left,anchor=west] at (1.2,2.1) {emergent string};
        \draw[purered,fill=purered] (1,1.6) circle (3pt);
        \node[align=left,anchor=west] at (1.2,1.6) {decompactification};
        \node at (2.7,-2.4) {D1-limit};
        \node at (2.7,-.3) {$\textrm{F1}_\textrm{B}$-limit};
        \node at (-2.3,2.5) {M2-limit};
        \node at (-2.1,.4) {M5-limit};
        \node at (-2.1,-2.4) {D4-limit};
        \node at (-2.1,-.3) {$\textrm{F1}_\textrm{A}$-limit};
        \node at (2.25,-1.25) {\scriptsize S-dual};
        \node at (-1.6,-1.25) {\scriptsize S-dual};
        \node at (0.35,-.85) {\scriptsize mirror};
        \node at (0.35,-1.3) {\scriptsize dual};
    \end{tikzpicture}
\caption{Under the identification of the 5D M-theory and 4D type IIA hypermultiplet moduli spaces, the quantum corrected emergent string limits of \cite{Marchesano:2019ifh,Baume:2019sry} correspond to decompactification limits in M-theory. \label{Fig_modulispaces}}
\end{figure}
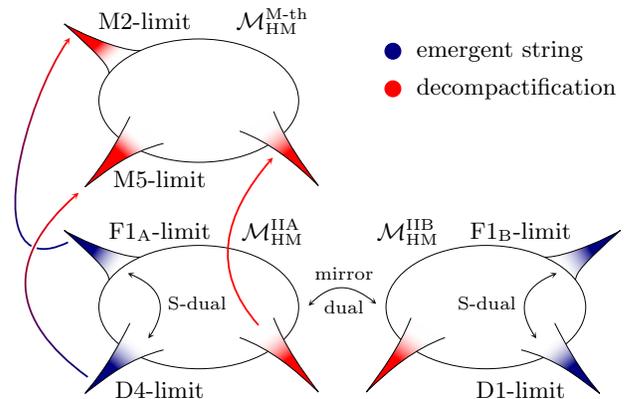

To provide a realisation of this quantum obstruction to membrane limits we exploit a useful fact: The hypermultiplet moduli space of M-theory on $\mathbb{R}^{1,4} \times X$ is identical to that of type IIA string theory on $\mathbb{R}^{1,3} \times X$. This implies, after mirror symmetry, an identification of the emergent string limits of \cite{Marchesano:2019ifh,Baume:2019sry} with the trajectories of the membrane limits under consideration, allowing us to translate the effect of the quantum corrections from one setting to the other. This leads to the conclusion that the putative membrane limits are quantum obstructed and the five-dimensional M-theory undergoes a decompactification limit instead. The situation is depicted in Figure \ref{Fig_modulispaces}.

The structure of the paper is the following. In Section \ref{sec:dimensional_reduction} we discuss how consistency under dimensional reduction for the Emergent String Conjecture forbids the existence of limits in which a ``critical'' membrane becomes light at the same rate as the KK scale. Section \ref{sec:cl-membrane-lims-M-th} gives a review of the hypermultiplet moduli space of five-dimensional M-theory and analyses the type of scalings needed in said moduli space to give rise to a classical membrane limit. In Section \ref{sec:type-II-results} we review the type IIB hypermultiplet moduli space limits of \cite{Marchesano:2019ifh,Baume:2019sry}. In Section \ref{sec_IIAandM} we translate these limits to the type IIA setting via mirror symmetry and make contact with the five-dimensional M-theory limits through the identification of the corresponding hypermultiplet moduli spaces. Section \ref{sec:quantum-obstruction} analyses how quantum corrections obstruct the classical membrane limits and instead lead to a decompactification of the theory. We conclude with a summary in Section \ref{sec:conclusions}.

\section{\label{sec:dimensional_reduction}Consistency under dimensional reduction}

Consistency under dimensional reduction has served  as a fruitful guiding principle for formulating swampland conjectures, most prominently in the context of the Weak Gravity Conjecture~\cite{Heidenreich:2015nta,Heidenreich:2015wga,Heidenreich:2016aqi,Andriolo:2018lvp,Palti:2019pca,Heidenreich:2019zkl,Cremonini:2020smy,Rudelius:2021oaz}. Here we would like to address the following question:
\begin{center}
{\bfseries Is the Emergent String Conjecture consistent under dimensional reduction?}
\end{center}

Suppose we have a theory in D dimensions and fix an infinite distance limit in its moduli space. Denote by $M_\textrm{KK}^{(\textrm{D})}$ the mass scale of the lightest KK tower in the limit and assume that $M_\textrm{KK}^{(\textrm{D})}/M_{\rm Pl}^{({\rm D})} \to 0$ asymptotically. Denote furthermore by $T_\textrm{str}^{(\textrm{D})}$ the tension of the lightest {\it critical} string in the limit. The Emergent String Conjecture predicts that $T_\textrm{str}^{(\textrm{D})}\succsim \big(M_\textrm{KK}^{(\textrm{D})}\big)^2$ asymptotically. Under compactification on a circle of constant radius, as measured in units of the D-dimensional Planck scale, this relation will be preserved. In addition, the string will generate a tower of winding states on the circle. These will however not lead to any inconsistency.

The situation is different if, instead of a critical string, we consider a membrane with a (2+1)-dimensional worldvolume in the D-dimensional theory. Let us denote the tension of the lightest such brane by $T_\textrm{brane}^{(\textrm{D})}$. Assuming the existence of such a brane in the D-dimensional theory, its infinite distance limits will be characterised by the dimensionless parameter 
\begin{equation}
\mu=\frac{T_\textrm{brane}^{(\textrm{D})}}{\big(M_\textrm{KK}^{(\textrm{D})}\big)^3} \,.
\end{equation}
After compactification, the membrane will lead to a new string of tension $T_\textrm{str}^{(\textrm{D}-1)}=R_{S^1}\cdot T_\textrm{brane}^{(\textrm{D})}$ from wrapping the brane on the $S^1$ of radius $R_{S^1}$. In the sequel it is always understood that the membrane under consideration has the special property that the string obtained in this way is a critical string. We will sometimes call such objects {\it critical membranes}.

The special case $\mu=\textrm{const}.$ could be called an ``emergent membrane" limit. The reader will object that, unlike a critical string, a membrane is not expected to give rise to an infinite tower of particle-like excitations and hence the special nature of such a regime might be dubious. Thus, such a limit should be classified as a decompactification limit in view of the Emergent String Conjecture. We will see, however, that the existence of an emergent membrane limit can potentially lead to trouble with the Emergent String Conjecture in lower dimensions.

In the dimensionally reduced theory, the Emergent String Conjecture requires that $T_\textrm{str}^{(\textrm{D}-1)}\succsim \big(M_\textrm{KK}^{(\textrm{D-1})}\big)^2$. For a circle of constant radius of order one in units of $M_{\rm Pl}^{({\rm D})}$, $M_\textrm{KK}^{(\textrm{D-1})} \sim M_\textrm{KK}^{(\textrm{D})}$. This gives rise to a non-trivial constraint on admissible limits in the D–dimensional theory:\footnote{Clearly, one can also consider limits in which the circle radius scales with respect to  $M_{\rm Pl}^{({\rm D})}$, but the fact that we run into a constraint for the current limit is already enough for our argument.}
\begin{equation}
\label{eq:em_str_dim_red}
    \begin{gathered}
    \frac{T_\textrm{str}^{(\textrm{D}-1)}}{\big(M_\textrm{KK}^{(\textrm{D}-1)}\big)^2}
    \sim \frac{T_\textrm{brane}^{(\textrm{D})}\cdot R_{S^1}}{\big(M_\textrm{KK}^{(\textrm{D})}\big)^2}
    =\frac{T_\textrm{brane}^{(\textrm{D})}}{\big(M_{\textrm {KK}}^{(\textrm{D})}\big)^3}\cdot\frac{M_{\textrm{ KK}}^{(\textrm{D})}}{M_{{\textrm{KK}}}^{S^1}}\\
    \sim\mu\cdot\frac{M_{\textrm{KK}}^{(\textrm{D})}}{M_{{\textrm{KK}}}^{S^1}}
    =\mu\cdot\frac{M_{\textrm{KK}}^{(\textrm{D})}}{M_{{\textrm{Pl}}}^{(\textrm{D})}}\cdot \frac{M_{\textrm{Pl}}^{(\textrm{D})}}{M_{{\textrm{KK}}}^{S^1}}
    \sim \mu\cdot\frac{M_{\textrm{KK}}^{(\textrm{D})}}{M_{{\textrm{Pl}}}^{(\textrm{D})}}\stackrel{!}{\succsim} 1 \,.
    \end{gathered}
\end{equation}
Since the KK scale is assumed to approach zero in the D-dimensional theory, it is necessary that $\mu\to\infty$ in the limit. Hence, we find that the Emergent String Conjecture predicts a censorship of infinite distance limits in a higher-dimensional theory where a critical membrane sits at the same scale as (or below) the KK tower.

In the saturated case of an emergent string limit in the dimensionally reduced theory, $T_\textrm{str}^{(\textrm{D}-1)}\sim \big(M_\textrm{KK}^{(\textrm{D})}\big)^2$, we can extract the following scaling in the D-dimensional theory from~\eqref{eq:em_str_dim_red}:
\begin{equation}
    \left(\frac{M_\textrm{KK}^{(\textrm{D})}}{M_\textrm{Pl}^{(\textrm{D})}}\right)^3\sim \frac{1}{\mu^3}\;,\qquad 
    \frac{T_\textrm{brane}^{(\textrm{D})}}{\big(M_\textrm{Pl}^{(\textrm{D})}\big)^3}\sim \frac{1}{\mu^2}\;.
\label{eq:membrane-KK-ratio}
\end{equation}
In the remainder of this article, we will explore how this censorship is realized quantitatively in the five-dimensional setting of M-theory compactified on a Calabi-Yau threefold.

\section{Classical membrane limits in M-theory}
\label{sec:cl-membrane-lims-M-th}

In this section we briefly review the moduli space of M-theory compactified on a Calabi-Yau threefold and we analyse the conditions that need to be met to engineer classical membrane limits.

\subsection{The moduli space of M-theory on Calabi-Yau threefolds}
\label{sec:M-hypermultiplets}

M-theory compactified on a Calabi-Yau threefold gives rise to a five-dimensional $\mathcal{N} = 2$ supergravity theory containing a number of hypermultiplets and vector multiplets whose scalar fields parametrise the deformations of the internal space. Such a dimensional reduction of eleven-dimensional supergravity was carried out in \cite{Cadavid:1995bk}.

In $\textrm{D}=11$ the bosonic field content is given by the metric $G_{\hat{\mu}\hat{\nu}}$ and a 3-form gauge field $C_{{\hat{\mu}\hat{\nu}\hat{\rho}}}$. Upon dimensional reduction on a Calabi-Yau threefold $X$ with Hodge numbers $(h^{1,1},h^{2,1})$ we obtain $h^{1,1}-1$ vector multiplets and $h^{2,1}+1$ hypermultiplets. In the following we denote the splitting of indices by $\hat{\mu} = (\mu,i,\overline{i})$ ($\mu = 1, \dots, 5$, $i,\overline{i} = 1,2,3$).

The scalar components of the vector multiplets parametrising the moduli space are accounted for by the (real) scalars $G_{i\overline{j}}$ except for the overall volume of $X$, which in five-dimensional M-theory decouples from the vector multiplets. If we denote by $M_{\Lambda}$ the K\"ahler coordinates in the decomposition of the K\"ahler form, the vector multiplet moduli space coordinates would be given by $t_{\Lambda} = M_{\Lambda}/\mathcal{V}^{\frac{1}{3}}$ subject to the constraint $\mathcal{V}(t_{\Lambda}) = 1$, i.e.\ the vector multiplet moduli space of five-dimensional M-theory is a hypersurface cut from the real projection of that of type IIA string theory.

The volume scalar $\mathcal{V}$, the real scalar $C_{\mu\nu\rho}$ and the complex scalar $C_{ijk} = \epsilon_{ijk}D$ conform the four scalar degrees of freedom of the universal hypermultiplet $(\mathcal{V},C_{\mu\nu\rho},D)$. The remaining $h^{2,1}$ hypermultiplets are accounted for by the complex scalars $(G_{ij},C_{ij\overline{k}})$.

\subsection{Classical membrane limits}
\label{sec:cl-membrane-lims}

We would like to investigate if it is possible to engineer infinite distance limits in the hypermultiplet moduli space of five-dimensional M-theory along which a critical membrane in the sense defined in Section \ref{sec:dimensional_reduction} becomes exponentially light at the same rate as the KK scale. In analogy with \cite{Lee:2019wij,Klaewer:2020lfg} we would call them emergent membrane limits.

We will consider trajectories involving only the volume scalar $\mathcal{V}$ and the complex structure moduli $z^{a}$, $a = 1, \dots, h^{2,1}$, of $X$. Varying $z^{a}$ will affect the volume of certain 3-cycles. We denote the 3-cycles with the slowest and fastest scaling along the limit by $\mathcal{A}$ and $\mathcal{B}$, respectively. Then, the mass scales of the membranes and KK modes in the theory compare to the five-dimensional Planck scale as
\begin{align}
    \frac{T_{\textrm{M2}}}{\left( \MPlfive \right)^{3}} &\sim \frac{1}{\Veleven} \sim \left( \frac{T_{\textrm{M5}}}{\left( \MPlfive \right)^{6}} \right)^{\frac{1}{2}}\,,\\
    \frac{T_{\textrm{M5}|_{\mathcal{A}}}}{\left( \MPlfive \right)^{3}} &\sim \vol[\textrm{11D}]{\mathcal{A}} \frac{1}{\Veleven}\,,\\
    \left( \frac{M_{\textrm{KK}}}{\MPlfive} \right)^{3} &\sim \min{\left( \frac{1}{\vol[\textrm{11D}]{\mathcal{B}}}, \left( \frac{1}{\Veleven} \right)^{\frac{1}{2}} \right)} \frac{1}{\Veleven}
    \label{eq:KK_scale}\,,
\end{align}
where all the volumes are measured in eleven-dimensional units. At the classical level, we can distinguish two scenarios:
\begin{itemize}
    \item {\bf M2-limit}: The unwrapped M2-brane becomes light at the same rate as the KK modes, which happens if\footnote{This condition follows straightforwardly from~\eqref{eq:KK_scale} if the minimal KK scale comes from $\mathcal{B}$. If the minimal KK scale instead comes from $X$, it follows that $\Veleven\sim 1$ and hence $\vol[\textrm{11D}]{\mathcal{B}}\precsim 1$. If $\mathcal{B}$ is shrinking, so must be all other 3-cycles since by assumption $\mathcal{B}$ has the fastest scaling. In order to avoid a contradiction with $\Veleven\sim 1$, we should therefore again require $\vol[\textrm{11D}]{\mathcal{B}}\sim 1$.}
    \begin{equation}
        \vol[\textrm{11D}]{\mathcal{B}} \sim \textrm{constant}\,.
    \label{eq:M2-condition}
    \end{equation}
    \item {\bf M5-limit}: The M5-brane wrapped on the $\mathcal{A}$-cycle becomes light at the same rate as the KK modes, which requires that
    \begin{equation}
        \vol[\textrm{11D}]{\mathcal{A}} \sim \frac{1}{\vol[\textrm{11D}]{\mathcal{B}}}\,\quad   {\rm or} \quad  \vol[\textrm{11D}]{\mathcal{A}} \sim \left( \frac{1}{\Veleven} \right)^{\frac{1}{2}}\,.
    \label{eq:M5-condition}
    \end{equation}
\end{itemize}
In order for the membrane obtained by wrapping the M5-brane on the $\mathcal{A}$-cycle to correspond to a critical membrane (i.e.\ one which gives rise to a critical string once wrapped on a circle), the  $\mathcal{A}$-cycle must be a special Lagrangian three-torus \cite{Lambert:2019khh}. This will become particularly clear from the perspective of the dual type IIA string theory in Section \ref{sec_IIAandM}.

Obtaining membrane limits is therefore classically possible. However, the hypermultiplet moduli space of five-dimensional M-theory receives quantum corrections that could deflect the chosen trajectories, modifying their properties such that the membrane no longer becomes light at the same rate as the KK modes.

A similar phenomenon was observed for infinite distance limits in the hypermultiplet moduli space of type IIB string theory in \cite{Marchesano:2019ifh,Baume:2019sry}. There, the D1-limit presented an infinite distance trajectory along which classically a D1-string became tensionless faster than the KK scale. Such behaviour is expected to be forbidden after taking quantum corrections into account, as otherwise we would be able to decouple both scales and obtain critical four-dimensional strings with an infinite number of oscillation modes. Indeed, the authors found that including the relevant effects due to D($-$1)- and D1-instantons modified the trajectory in such a way that both scales became light at the same rate, yielding an emergent string limit. Similar quantum obstructions to pathological string limits in the vector multiplet moduli space of type II theories were analysed in \cite{Lee:2019wij} and for 4D $\mathcal{N} = 1$ theories in \cite{Klaewer:2020lfg}.

Given the above, we need a way to take into account the quantum corrections to the hypermultiplet moduli space of five-dimensional M-theory. Our strategy will be to first study the problem in the hypermultiplet moduli space of type IIA string theory and to then translate the results to the M-theory setup.

\section{\label{RelationToIIA} Review: Type IIB hypermultiplet limits}
\label{sec:type-II-results}

The quantum corrections to the hypermultiplet moduli space of type II string theories compactified on Calabi-Yau threefolds have been extensively studied in the literature; see \cite{Alexandrov:2011va,Alexandrov:2013yva} and references therein. In the settings in which only mutually local D-instantons contribute, the quantum-corrected metric for the hypermultiplet moduli space can be given explicitly.\footnote{This metric was first computed in~\cite{Alexandrov:2014sya}. See also \cite{Cortes:2021vdm} for a recent mathematical treatment of quaternionic Kähler metrics including the case of the hypermultiplet metric with mutually local D-instanton corrections.} This fact was exploited in \cite{Marchesano:2019ifh,Baume:2019sry}, where all the limits fall in this category.

We are investigating limits in the hypermultiplet moduli space of M-theory on $\mathbb{R}^{1,4} \times X$. Since the radius of the M-theory circle, measured in five-dimensional Planck units, sits in a vector multiplet from the point of view of the dimensionally reduced theory, this moduli space coincides with the hypermultiplet moduli space of type IIA string theory on $\mathbb{R}^{1,3} \times X$ \cite{Pioline:2009qt}. By then exploiting mirror symmetry to type IIB on $\mathbb{R}^{1,3} \times Y$ we will be able to translate the quantum corrections computed in \cite{Marchesano:2019ifh,Baume:2019sry} to the membrane limits under consideration.

We would like to stress that the relation between the limits is at the level of the trajectories in the respective hypermultiplet moduli spaces, once viewed as hypermultiplets of the four-dimensional type IIA compactification on $X$ and once as those of the five-dimensional M-theory compactification. In the first case, the trajectories describe emergent string limits, while  the nature of the limit in the second case is the subject of our investigation. We will elaborate on this point more in Section \ref{subsec_Id}.

In this section we review the results of \cite{Marchesano:2019ifh,Baume:2019sry} on emergent string limits in the hypermultiplet moduli space of type IIB string theory and the quantum corrections they receive. At the end we also comment on the reasons behind focusing on the class of limits under consideration here.

\subsection{Type II hypermultiplet moduli spaces}

To fix notation, let us list the scalars parametrising the hypermultiplet moduli space of type II string theories. Excellent reviews on said moduli spaces and the quantum corrections that they receive can be found in \cite{Alexandrov:2011va,Alexandrov:2013yva}.

For type IIB string theory on a Calabi-Yau threefold $Y$ we have $h^{1,1}(Y) + 1$ hypermultiplets with the following bosonic content:
\begin{align}
    \textrm{universal hypermultiplet: } &(\tau_{\IIB},b^{0},c^{0})\,,\\
    h^{1,1}(Y) \textrm{ hypermultiplets: } &(z^{a}_{\IIB},c^{a},d^{a})\,.
\end{align}
Here $\tau_{\IIB} = C_{0} + ie^{-\phi_{\IIB}} = \tau_{1}^{\IIB} + i \tau_{2}^{\IIB}$ is the ten-dimensional type IIB axio-dilaton and $b^{0}$ and $c^{0}$ are the axions dual to the four-dimensional components of the $B_{2}$ and $C_{2}$ 2-forms respectively. Furthermore, $z^{a}_{\IIB} = b^{a} + i t^{a}$ are the complexified K\"ahler moduli for the decomposition of the K\"ahler form over a basis $\{\gamma^{a}_{\IIB}\}$ of 2-cycles and $c^{a}$ and $d^{a}$ are two axions related to the integrals of the $C_{2}$ and $C_{4}$ forms over the same basis of 2-cycles. 

Defining the four-dimensional dilaton by
\begin{equation}
    \left( \MPlfour \right)^{2} = 4\pi e^{-2\varphi_{4}} M_{s}^{2} = 2\pi (\tau_{2}^{\IIB})^{2} \mathcal{V}_{Y}(t^{a}) M_{s}^{2}
\end{equation}
we have appropriate coordinates to express the classical hypermultiplet moduli space metric as
\begin{equation}
    ds^{2}_{\mathcal{M}_{\textrm{HM}}^{\IIB}} = \frac{1}{2} \left( \varphi_{4} \right)^{2} + g_{a\overline{b}} dz^{a}_{\IIB}d\overline{z}^{b}_{\IIB} + (\textrm{axions})\,.
\label{eq:hm-metric-cl}
\end{equation}
Corresponding to the membrane limits with which we would like to make contact, and as in the trajectories considered in \cite{Marchesano:2019ifh,Baume:2019sry}, we will keep the axions set to zero.

Type IIA string theory compactified on a Calabi-Yau threefold $X$ presents $h^{2,1}(X)+1$ hypermultiplets, whose bosonic content is as follows:
\begin{align}
    \textrm{universal hypermultiplet: } &(\phi_{\IIA},\sigma,\zeta^{0},\tilde{\zeta}_{0})\,,\\
    h^{2,1}(X) \textrm{ hypermultiplets: } &(z^{a}_{\IIA},\zeta^{a},\tilde{\zeta}_{a})\,.
\end{align}
Here $\phi_{\IIA}$ is the ten-dimensional type IIA dilaton, $\sigma$ is the Neveu-Schwarz axion dual to the four-dimensional 2-form $B_{2}$ and $\zeta^{0}$ and $\tilde{\zeta}_{0}$ are obtained by integrating the 3-form $C_{3}$ over the 3-cycle $\gamma^{0}$ dual to the unique holomorphic $(3,0)$-form $\Omega_{X}$ of $X$ and its symplectic pair $\gamma_{0}$ respectively. The remaining $h^{2,1}(X)$ hypermultiplets involve the complex structure moduli $z^{a}_{\IIA} = X^{a}/X^{0}$, where $(X^{0},X^{a})$ are the $\Omega_{X}$ periods of $X$. Finally, $\zeta^{a}$ and $\tilde{\zeta}_{a}$ are obtained by integrating $C_{3}$ over the A- and B-cycles conforming the $\{\gamma^{a}_{\IIA}, \gamma_{a}^{\IIA}\}$ basis of (1,2)- and (2,1)-cycles. 

The four-dimensional dilaton is defined by
\begin{equation}
    \left( \MPlfour \right)^{2} = 4\pi e^{-2\varphi_{4}} M_{s}^{2} = \pi \mathcal{R}^{2} K(z_{\IIA}^{a},\overline{z}_{\IIA}^{a}) M_{s}^{2}\,,
\end{equation}
where $\mathcal{R}$ is related to the ten-dimensional type IIA string coupling as displayed in \eqref{eq:mirror-map}. The classical hypermultiplet moduli space metric is \textit{mutatis mutandis} the same as \eqref{eq:hm-metric-cl}. Again, the axions will be set to zero in the following.
Throughout the text we will denote the string coupling by $g_{\textrm{IIA(B)}} = 1/\tau_2^{\textrm{IIA(B)}}$.

\subsection{Classical string limits}

Out of the limits in the hypermultiplet moduli space of type IIB string theory discussed in \cite{Marchesano:2019ifh,Baume:2019sry} we are interested in recalling the properties of the D1- and F1-limits, which we will later identify with the M5- and M2-limits, respectively.

\vspace{\baselineskip}
\textbf{D1-limit:} The classical D1-limit corresponds to a strong coupling, large volume limit in which the 2-cycles of $Y$ are uniformly scaled.\footnote{See Section \ref{sec_otherlimits} for a discussion on the systematics behind the infinite distance limits.} More concretely, the scaling along the trajectory is given by
\begin{equation}
    \textrm{D1:}\quad g_{\IIB} \sim \lambda^{\frac{3}{2}}\,,\quad t^{a} \sim \lambda\,,\quad \lambda \rightarrow \infty\,.
\end{equation}
The object becoming massless at the fastest rate, in the classical limit, is the D1-string with tension
\begin{equation}
    \frac{T_{\textrm{D1}}}{\left( \MPlfour \right)^{2}} = \frac{1}{g_{\IIB}} \left( \frac{M_{s}}{\MPlfour} \right)^{2} \sim \frac{1}{\lambda^{\frac{3}{2}}}\,.
\end{equation}
The KK scale is set by the overall volume of the manifold $Y$ as
\begin{equation}
    \left( \frac{M_{\textrm{KK}}}{\MPlfour} \right)^{2} = \frac{1}{\mathcal{V}_{Y}^{\frac{1}{3}}} \left( \frac{M_{s}}{\MPlfour} \right)^{2} \sim \frac{1}{\lambda}\,.
\end{equation}

\vspace{\baselineskip}
\textbf{F1-limit:} The F1-limit is the S-dual of the D1-limit. Taking into account the action of S-duality,
\begin{equation}
    \tau'_{\IIB} = -\frac{1}{\tau_{\IIB}}\,,\quad t^{'a} = |\tau_{\IIB}| t^{a}\,,\quad e^{-2\varphi'_{4}} = g_{\IIB} e^{-2\varphi_{4}}\,,
\end{equation}
we obtain the trajectory of the weak coupling, small volume limit
\begin{equation}
    \textrm{F1:}\quad g'_{\IIB} \sim \frac{1}{\lambda^{\frac{3}{2}}}\,,\quad t^{'a} \sim \frac{1}{\lambda^{\frac{1}{2}}}\,,\quad \lambda \rightarrow \infty\,.
\end{equation}
As expected from S-duality, it is now the F1-string that classically becomes the lightest,
\begin{equation}
    \frac{T_{\textrm{F1}}}{\left( \MPlfour \right)^{2}} = \left( \frac{M'_{s}}{\MPlfour} \right)^{2} \sim \frac{1}{\lambda^{\frac{3}{2}}}\,,
\end{equation}
with the KK scale falling once again behind as
\begin{equation}
    \left( \frac{M_{\textrm{KK}}}{\MPlfour} \right)^{2} = \frac{1}{\mathcal{V}_{Y}^{'\frac{1}{3}}} \left( \frac{M'_{s}}{\MPlfour} \right)^{2} \sim \frac{1}{\lambda}\,.
\end{equation}
Both of these classical limits, in which a critical string becomes parametrically lighter than the KK scale, get deflected by quantum corrections \cite{Marchesano:2019ifh,Baume:2019sry}.

\subsection{Quantum corrections}

Along the F1-limit, the relevant quantum corrections are the $\alpha'$-corrections and the worldsheet instanton contributions. For the D1-limit it is the D(-1)- and D1-instantons that will give the relevant corrections, which constitute a mutually local configuration of instantons. The two sets of quantum corrections are S-dual to each other, as one would have expected.

When only D($-$1)- and D1-instantons contribute, enough continuous shift symmetries in the Ramond-Ramond sector remain unbroken so that the hypermultiplet moduli space simplifies and can be described in terms of tensor multiplets. The quantum corrections can then be captured in an object known as the contact potential $\chi$, which is a quantum-corrected version of the four-dimensional dilaton,
\begin{equation}
    \left( \frac{M_{s}}{\MPlfour} \right)^{2} \sim e^{2\varphi_{4}} \xrightarrow{\QC} \frac{1}{\chi}\,,\quad \chi' = \frac{\chi}{|\tau_{\IIB}|} \xleftrightarrow{\textrm{S-dual}} \chi\,.
\end{equation}
It was argued in~\cite{Marchesano:2019ifh} that this quantity determines a metric that asymptotically approximates the one on the hypermultiplet space in the relevant limits. In particular, it acts as an approximate K\"ahler potential
\begin{equation}
    K = -\log \chi
\end{equation}
for the metric $g_{a\overline{b}}$ on the space of complexified K\"ahler moduli.
The explicit expression for $\chi$ was found in \cite{Robles-Llana:2006hby} by exploiting the c-map and considerations of $\textrm{SL}(2,\mathbb{Z})$ invariance, and recast in a new form by Poisson resummation in \cite{Robles-Llana:2007bbv} to make the role of the D($-$1)- and D1-instantons manifest.

This was used in \cite{Marchesano:2019ifh,Baume:2019sry} to analyse exactly which contributions need to be taken into account along the F1- and D1-limits, and how they modify the classical trajectory. The pathological behaviour classically found for the string limits is thereby removed.

\vspace{\baselineskip}
\textbf{F1-limit:} For the F1-limit, in which classically the volume of the manifold shrinks to zero, it was argued in \cite{Baume:2019sry} that instead one eventually reaches a minimal quantum-corrected volume such that the K\"ahler coordinates freeze but one still encounters an infinite distance limit along the quantum-corrected four-dimensional dilaton. This was supported by analysing a similar trajectory in the vector multiplet moduli space of type IIA string theory on $Y$, which was then embedded into the hypermultiplet moduli space of type IIB string theory on $Y$ via the c-map. The effect of this is that the contribution of the worldsheet instantons freezes deep enough into the trajectory, i.e.\ once the quantum volume has been reached, the quantum-corrected quantities contain a constant piece corresponding to the worldsheet instantons that no longer affects the functional dependence on the parameter $\lambda$. The scaling of the four-dimensional dilaton is quantum corrected to
\begin{equation}
    \left( \frac{M'_{s}}{\MPlfour} \right)^{2} \sim \frac{1}{\lambda^{\frac{3}{2}}} \xrightarrow{\QC} \frac{1}{\chi'} \sim \frac{1}{\lambda^{3}}\,,
\label{eq:4D-dilaton-F1}
\end{equation}
which results in the F1-string tension
\begin{equation}
    \frac{T_{\textrm{F1}}}{\left( \MPlfour \right)^{2}} = \frac{1}{\chi'} \sim \frac{1}{\lambda^{3}}
\end{equation}
and the KK scale
\begin{equation}
    \left( \frac{M_{\textrm{KK}}}{\MPlfour} \right)^{2} = \frac{1}{\left( \mathcal{V}_{Y}^{'\QC} \right)^{\frac{1}{3}}} \frac{1}{\chi'} \sim \frac{1}{\lambda^{3}}
\end{equation}
decreasing at the same rate. In other words, the quantum corrections deflect the pathological string limit in the necessary way to obtain an emergent string limit.

\vspace{\baselineskip}
\textbf{D1-limit:} The results for the F1-limit were then translated to the D1-limit setting by S-duality. Under S-duality, freezing of the worldsheet instantons is translated into freezing of the D1-instanton contributions deep enough along the trajectory. For a classical D1-type trajectory the relative scaling of the K\"ahler coordinates and the string coupling is such that
\begin{equation}
    \frac{t^{a}}{g_{\IIB}} \xrightarrow{\lambda \rightarrow \infty} 0
\end{equation}
(note that if the quotient goes to $\infty$ we have a decompactification limit, see below). This corresponds under S-duality to an F1-type limit in which the classical volume of the manifold is shrinking. Now, in order for the contribution of the D1-instantons to freeze, we need the scaling of the K\"ahler moduli to get quantum accelerated deep enough along the trajectory to obtain
\begin{equation}
    \frac{t^{a}_{\QC}}{g_{\IIB}} \xrightarrow{\lambda \rightarrow \infty} 1\,.
\label{eq:quantum-scaling}
\end{equation}
This yields a quantum-corrected trajectory for the four-dimensional dilaton
\begin{equation}
    \left( \frac{M_{s}}{\MPlfour} \right)^{2} \sim 1 \xrightarrow{\QC} \frac{1}{\chi} \sim \frac{1}{\lambda^{\frac{3}{2}}}\,,
\label{eq:4D-dilaton-D1}
\end{equation}
which is compatible with \eqref{eq:4D-dilaton-F1} under S-duality. As a result, the D1-string tension,
\begin{equation}
    \frac{T_{\textrm{D1}}}{\left( \MPlfour \right)^{2}} = \frac{1}{g_{\IIB}} \frac{1}{\chi} \sim \frac{1}{\lambda^{3}}\,,
\end{equation}
and the KK mass,
\begin{equation}
    \left( \frac{M_{\textrm{KK}}}{\MPlfour} \right)^{2} = \frac{1}{\left( \mathcal{V}_{Y}^{\QC} \right)^{\frac{1}{3}}} \frac{1}{\chi} \sim \frac{1}{\lambda^{3}}\,,
\end{equation}
scale in the same way. Once again, the quantum corrections modify the classical trajectory so as to remove the pathological behaviour and to yield an emergent string limit.

\subsection{Other possible limits} \label{sec_otherlimits}

We have just reviewed the physics of the D1-limit and its S-dual, the F1-limit. Throughout the text we will only consider limits that are related, after an appropriate chain of identifications, to these two limits. This restriction, present already in \cite{Marchesano:2019ifh,Baume:2019sry}, stems from the inherent difficulty of performing explicit computations for the quantum corrections to trajectories in the hypermultiplet moduli space of type II string theories. Therefore, we have to content ourselves with configurations in which only mutually local D-instantons contribute, for which the problem is tractable.

Consider a homogeneously scaling manifold with
\begin{equation}
    t^{a} \sim \lambda^{\mu}\,,\quad g_{\IIB} \sim \lambda^{\nu}\,.
\end{equation}
As long as we are in the classical large volume regime, the non-perturbative contributions then scale like
\begin{align}
    S_{\textrm{WS}} &\sim {\cal V}_{\textrm{2-cycle}} \sim t^{a} \sim \lambda^{\mu}\,,\\
    S_{\textrm{D(-1)}} &\sim \frac{{\cal V}_{\textrm{0-cycle}}}{g_{\IIB}} \sim \frac{1}{g_{\IIB}} \sim \lambda^{-\nu}\,,\\
    S_{\textrm{D1}} &\sim \frac{{\cal V}_{\textrm{2-cycle}}}{g_{\IIB}} \sim \frac{t^{a}}{g_{\IIB}} \sim \lambda^{\mu-\nu}\,,\\
    S_{\textrm{D3}} &\sim \frac{{\cal V}_{\textrm{4-cycle}}}{g_{\IIB}} \sim \frac{(t^{a})^{2}}{g_{\IIB}} \sim \lambda^{2\mu-\nu}\,,\\
    S_{\textrm{D5}} &\sim \frac{{\cal V}_{\textrm{6-cycle}}}{g_{\IIB}} \sim \frac{(t^{a})^{3}}{g_{\IIB}} \sim \lambda^{3\mu-\nu}\,,\\
    S_{\textrm{NS5}} &\sim \frac{{\cal V}_{\textrm{6-cycle}}}{g_{\IIB}^{2}} \sim \frac{(t^{a})^{3}}{g_{\IIB}^{2}} \sim \lambda^{3\mu-2\nu}\,.
\end{align}
Demanding that D3-, D5- and NS5-instantons do not give relevant contributions along the limit\footnote{Mutually local instanton configurations in which D3- and D5-instantons contribute do exist. However, after mirror symmetry to the type IIA side, a symplectic transformation and mirror symmetry back to the IIB side, they can be written as a configuration in which the only D-instantons whose action decreases are the D($-$1)- and D1-instantons. We therefore consider only this mutually local type of configuration.} constrains $\mu$ and $\nu$ to either fulfil the inequality
\begin{equation}
    \mu > 0\,,\quad \textrm{and}\quad \nu \leq \frac{3\mu}{2}
\label{eq:D1-type-ineq}
\end{equation}
or the inequality
\begin{equation}
    \mu \leq 0\,, \quad \textrm{and}\quad \nu \leq 3\mu\,.
\label{eq:F1-type-ineq}
\end{equation}
Trajectories fulfilling condition \eqref{eq:D1-type-ineq} are related by S-duality to trajectories that satisfy condition \eqref{eq:F1-type-ineq}, and vice versa. The D1-limit saturates condition \eqref{eq:D1-type-ineq} and, as a consequence, the F1-limit lies in the regime of condition \eqref{eq:F1-type-ineq}. For $\lambda \to \infty$ the latter of course leaves the regime of validity of the large volume approximation and quantum corrections to the volume formulae become important, as discussed.

Looking closer at \eqref{eq:D1-type-ineq} we see that it also includes trajectories that classically lead to decompactification limits and are therefore not of our interest. Computing the classical tension of the different object and imposing that the KK scale sits below all of them amounts to the condition $\mu > \nu$. To exclude the classical decompactification limits we therefore also impose $\mu \leq \nu$, obtaining
\begin{equation}
    0 < \mu \leq \nu \leq \frac{3\mu}{2}\,.
\end{equation}
The classical D1-limit trajectory saturates $\nu \leq 3\mu/2$, while the quantum-corrected trajectory, after taking \eqref{eq:quantum-scaling} into account, saturates $\mu \leq \nu$.

Within the class of limits in which the manifold is homogeneously growing or shrinking, this explains the choice of limits. We could ask ourselves if inhomogeneous limits could lead to other situations in which the problem is also tractable and that present different physics.

Beyond an overall scaling of the manifold, the large distance finite volume limits in the classical K\"ahler moduli space of a Calabi-Yau threefold were classified in \cite{Lee:2019wij}. There are three possibilities: Type $T^{2}$, K3 and $T^{4}$ limits. The nomenclature refers to the type of fibre that shrinks along the limit, given by a genus-one curve $T^{2}$, a K3 surface or an Abelian surface $T^{4}$, respectively.

A particular realisation of a Type $T^{2}$ limit was considered in \cite{Baume:2019sry}, termed D3-limit in reference to the leading tensionless string obtained from wrapping D3-branes on the shrinking elliptic fibre. The D3-limit is obtained as a Fourier-Mukai transform of the D1-limit. This means that, in spite of D3-instantons contributing to the D3-limit, the quantum corrections can still be extracted from the results for the D1-limit. It also means that the physics of the two limits is equivalent, and therefore there is no need to discuss the D3-limit separately.

The same discussion should in principle go through for a Type $T^{4}$ limit. After translating such a limit to a D1-limit via a Fourier-Mukai transform we can essentially repeat the analysis. Such limits are expected to be physically equivalent to those D1-limits with tractable instanton contributions, and we therefore do not pursue this direction further. By contrast, a type IIB 5-brane on a shrinking K3-fiber does not give rise to a critical string\footnote{For example, if the K3 fibre is also elliptically fibred, a Fourier-Mukai transform along the elliptic fibre transforms a D5-brane on the K3 fibre to a D3-brane wrapping the base of the K3, which yields a non-critical string. By S-duality similar conclusions hold for wrapped type IIB NS5-branes.} and hence such degenerations do not lead to an emergent string limit. This is a notable difference to the
emergent heterotic string limits in type IIA string theory/M-theory associated with a shrinking K3-fiber wrapped by an NS5/M5-brane \cite{Lee:2019wij}.

\section{Type IIA hypermultiplet limits and relation to M-theory}
\label{sec_IIAandM}

We now use mirror symmetry to translate the trajectories in the hypermultiplet moduli space of type IIB string theory studied in the previous section to limits in the type IIA setting. We also review the equivalence between the hypermultiplet moduli spaces necessary to make contact with five-dimensional M-theory.
This section provides the technical foundation for the analysis in Section \ref{sec:quantum-obstruction}.

\subsection{Mirror map}
\label{sec:mirror-map}

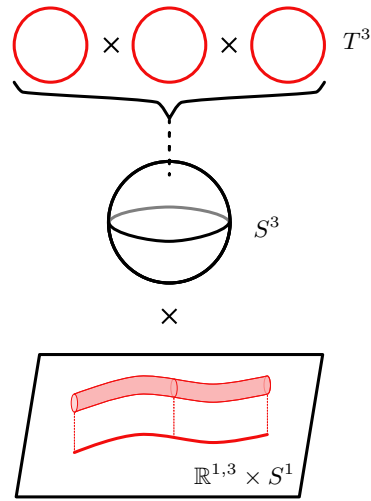
\begin{figure}[t!]
    \begin{tikzpicture}
        \node[inner sep=0pt] at (0,0) {\includegraphics[width=0.25\textwidth]{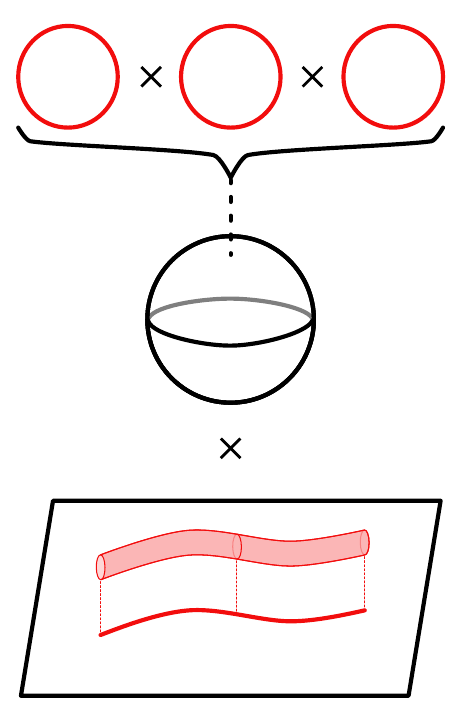}};
        \node at (2.5,2.8) {$T^3$};
        \node at (1.3,.3) {$S^3$};
        \node at (1,-3) {$\mathbb{R}^{1,3}\times S^1$};
    \end{tikzpicture}
\caption{The SYZ fibration structure taken into consideration for the limits in the region of large volume/LCS. \label{SYZ}}
\end{figure}

In the picture advocated by Strominger, Yau, and Zaslow (SYZ) \cite{Strominger:1996it}, mirror symmetry is understood, near the large complex structure point of $X$, as T-duality along a special Lagrangian 3-cycle with $T^{3}$ topology by which $X$ is fibred. We will refer to this cycle as the SYZ-cycle. See Figure \ref{SYZ} for an illustration.

The classical mirror map was found in \cite{Bohm:1999uk}, with the role of quantum corrections, such as instanton effects, discussed in \cite{Alexandrov:2009qq,Alexandrov:2012bu,Alexandrov:2013mha}. The quantum-corrected mirror map for the fields we are interested in coincides with the classical one for D-instanton configurations with vanishing magnetic charge \cite{Alexandrov:2014sya}, such as the ones we are considering. Therefore, it will suffice for our purposes to take into account the relations
\begin{equation}
    z^{a}_{\IIA} = z^{a}_{\IIB}\,,\quad \varphi_{4}^{\IIA} = \varphi_{4}^{\IIB} \Leftrightarrow \frac{g_{\IIA}}{\mathcal{V}_{\SYZ}^{\IIA}} = g_{\IIB} = \frac{1}{2\mathcal{R}}\,.
\label{eq:mirror-map}
\end{equation}
Here $\mathcal{V}_{\SYZ}^{\IIA}$ is the volume of the SYZ-cycle.

The D-instantons are accordingly mapped under mirror symmetry \cite{Robles-Llana:2007bbv}. D($-$1)-instantons wrapping a 0-cycle in $Y$ are mapped to D2-instantons wrapping $\gamma^{0}$ in $X$, which we can identify with the SYZ-cycle, i.e.
\begin{equation}
        \mathcal{V}_{\SYZ} = \mathcal{V}_{\gamma^{0}}\,.
\end{equation}
D1-instantons wrapping a 2-cycle $k_{a} \gamma^{a}_{\IIB}$ in $Y$ with $n$ units of D($-$1)-charge are mapped to D2-instantons wrapping a special Lagrangian 3-cycle $k_{a}\gamma^{a}_{\IIA} + n\gamma^{0}$ in $X$. Therefore, mirror symmetry identifies
\begin{align}
    \textrm{D(-1)}|_{\textrm{0-cycle}} &\longleftrightarrow \textrm{D2}|_{\gamma^{0} = \textrm{SYZ-cycle}}\,,\\
    \textrm{D1}|_{k_{a}\gamma^{a}_{\IIB}} &\longleftrightarrow \textrm{D2}|_{k_{a}\gamma^{a}_{\IIA}} \label{eqD1inst}\,.
\end{align}

Strings coming from D$(p+2)$-branes wrapping $(p+1)$-cycles and D$p$-instantons supported on the same cycles scale in the same way on both sides of the mirror map.

\vspace{\baselineskip}
\textbf{Volume of special Lagrangian 3-cycles:} On the type IIA side, knowing the scaling of the volume of some special Lagrangian 3-cycles will prove to be necessary in order to compute the tensions of the objects under scrutiny.

The volume of a special Lagrangian 3-cycle $\Gamma$ in $X$ is given by (see e.g.\ \cite{Lee:2019wij})
\begin{equation}
    \mathcal{V}_{\Gamma}^{\IIA} = \frac{(8 \Vfour)^{\frac{1}{2}}}{\left( i \int_{X} \Omega_{X} \wedge \overline{\Omega}_{X} \right)^{\frac{1}{2}}} \textrm{Im} \int_{\Gamma} e^{-i \theta} \Omega_{X}\,,
\label{eq:3-cycle-vol}
\end{equation}
where $\theta$ is related to the calibration and the superscript IIA indicates that the volume is measured in string units. The first thing we observe is that, since on the type IIA side we are taking a complex structure limit, $\Vfour = \int_{X} \left( J_{X}^{\IIA} \right)^{3}$ is constant. In order to make effective use of this formula we need to justify how the scaling of the rest of the quantities involved will be extracted via mirror symmetry from the type IIB trajectories.

We start by looking at the classical D1- and F1-limits. We have seen that the corresponding trajectories are modified by taking into account the appropriate quantum corrections relevant deep into the limit, but at the initial stages of the trajectory we are still in a regime of reasonably large volume and moderate string coupling. This situation corresponds, under mirror symmetry, to the large complex structure (LCS) region in complex structure moduli space. As long as this approximation is valid, the denominator scales like
\begin{equation}
    \left( i \int_{X} \Omega_{X} \wedge \overline{\Omega}_{X} \right)^{\frac{1}{2}} \sim \left( \int_{Y} J_{Y}^{3} \right)^{\frac{1}{2}} \sim (t^{a})^{\frac{3}{2}}\,.
\end{equation}
The period vector, evaluated for the 3-cycles in $\{\gamma^{0},\gamma^{a}_{\IIA},\gamma_{a}^{\IIA},\gamma_{0}\} = \{\gamma^{\alpha}\}$, scales as
\begin{equation}
    \textrm{Im} \int_{\gamma^{\alpha}} e^{-i \theta} \Omega_{X} \sim \Pi^{\alpha}(t^{a})\,,\quad \Pi^{\alpha}(t^{a}) \sim (1,t^{a},(t^{a})^{2},(t^{a})^{3})\,,
\end{equation}
where we have exploited the structure of the LCS periods.

Therefore, for the D1-limit the 3-cycle with the slowest scaling $\mathcal{V}_{\mathcal{A}}$ will be the SYZ-cycle with
\begin{equation}
    \mathcal{V}_{\SYZ}^{\IIA} \sim \frac{1}{(t^{a})^{\frac{3}{2}}}\,,
\label{eq:SYZ-vol}
\end{equation}
while its symplectic dual $\gamma_{0}$ will be the one with the fastest scaling $\mathcal{V}_{\mathcal{B}}$ with
\begin{equation}
    \mathcal{V}_{\gamma_{0}}^{\IIA} \sim (t^{a})^{\frac{3}{2}}\,.
\label{eq:SYZ-pair-vol}
\end{equation}
In the F1-limit, classically $t^{a} \rightarrow 0$ as $\lambda \rightarrow \infty$ and therefore the roles of the slowest and fastest scaling cycles are exchanged.

The above reasoning is valid only in an appropriate region of moduli space. By definition, the classical limits are obtained by taking the scaling of certain quantities beyond their regime of applicability, already on the type IIB side, i.e.\ ignoring the relevant quantum corrections. Therefore, it is implicit in the nomenclature that in order to obtain the classical type IIA trajectories we can use the above formulae for both the D1- and F1-limits.

Of course, we are interested in the quantum-corrected limits deep along the trajectory, which are modified compared to the classical situation in such a way that the pathological string limits found by naively extrapolating the classical scalings are removed. Here we should be more careful, as for example the behaviour of the periods could change as we move in moduli space.

The D1-limit is a large volume limit that should correspond via mirror symmetry to an LCS limit in complex structure moduli space. Therefore, we can continue to trust both the SYZ fibration structure and the scaling of the periods, which we used to extract the dependence of the volume of the relevant 3-cycles on the mirror dual $t^{a}$ variables. We then just need to take into consideration the quantum acceleration of the scaling discussed around \eqref{eq:quantum-scaling} in order to account for the quantum corrections.

In the F1-limit the situation is more subtle, as we are moving away from the large volume region. Indeed, the naive vanishing of the volume along the classical trajectory arises from taking the large volume/LCS analysis too far. This was argued in \cite{Baume:2019sry} by analysing the analogous problem in the vector multiplet moduli space and embedding the trajectory into the hypermultiplet moduli space via the c-map. If we denote by $y^{a}$ a set of coordinates in complex structure moduli space, the naive vanishing of the volume on the K\"ahler side would be found by employing the classical mirror map
\begin{equation}
    z^{a}_{\IIA} = z^{a}_{\IIB} = \frac{1}{2\pi i} \log(y^{a})\,,
\end{equation}
valid around the LCS point $y^{a} = 0$, also deep along the trajectory. The small volume point corresponding to $y^{a} \rightarrow 1$ is actually a constant volume point, as the periods, and consequently the quantum-corrected mirror map, tend to a constant value. For specific examples of this phenomenon in the vector multiplet moduli space see \cite{Blumenhagen:2018nts,Joshi:2019nzi}. In this way, the type IIB K\"ahler coordinates $z^{a}_{\IIB}$ freeze at late stages of the F1-limit and by mirror symmetry so do the type IIA complex structure coordinates $z^{a}_{\IIA}$. Since in the mirror dual to the F1-limit no D-instantons contribute significantly and the hypermultiplet moduli space is $\alpha'$-exact,\footnote{\label{fn:classical-quantum}From the point of view of the type IIA mirror of the F1-limit taking the complete form of the periods as opposed to only considering the leading terms around the LCS is not a quantum correction, as both situations are classical. Therefore, the term ``quantum corrections" for this limit refers to the splitting of the quantities that one would observe on the K\"ahler moduli space side.}, the fact that the $z^{a}_{\IIA}$ approach a constant value should also be seen directly from the structure of the periods in concrete examples. Ultimately, the important fact is that even when the formulae derived above stop being valid along the F1-limit and its mirror dual, deep enough along the trajectory we know that $z^{a}_{\IIA}$ will become constant. This is enough information to obtain the asymptotic behaviour for the mass scales appearing in the problem, and reproduces the results found after quantum corrections on the type IIB side.

After these preliminaries, let us now analyse the mirror duals of the D1- and the F1-limits.

\vspace{\baselineskip}
\textbf{D4-limit:} First, consider the D1-limit. We have tensionless D1-strings and D($-$1)-instantons wrapping a 0-cycle in $Y$. Under mirror symmetry the lightest object must hence be a D4-string wrapping the SYZ-cycle. Therefore, we call this limit the D4-limit.
Indeed, under the three T-dualities which define mirror symmetry in the SYZ picture, a D4-brane along the $T^3$ fibre maps to a D1-brane in type IIB.

The type IIA string coupling is given, taking into account \eqref{eq:mirror-map}, \eqref{eq:SYZ-vol} and \eqref{eq:quantum-scaling}, by
\begin{equation}
    g_{\IIA} \sim \frac{g_{\IIB}}{(t^{a})^{\frac{3}{2}}} \sim 1 \xrightarrow{\QC} \frac{1}{\lambda^{\frac{3}{4}}}\,.
\label{eq:D4-gIIA}
\end{equation}
The tension of the D4-brane wrapping the SYZ-cycle scales, using \eqref{eq:mirror-map} and \eqref{eq:4D-dilaton-D1}, like
\begin{equation}
    \frac{T_{\textrm{D4}|_{\SYZ}}}{\left( \MPlfour \right)^{2}} = \frac{\mathcal{V}_{\SYZ}^{\IIA}}{g_{\IIA}} \left( \frac{M_{s}}{\MPlfour} \right)^{2} \sim \frac{1}{\lambda^{\frac{3}{2}}} \xrightarrow{\QC} \frac{1}{\lambda^{3}}\,.
\end{equation}
In this complex structure limit the volume of $X$ measured in string units remains unaltered, but the manifold becomes highly inhomogeneous, with the KK scale set by the $\gamma_{0}$-cycle. From \eqref{eq:SYZ-pair-vol}, \eqref{eq:quantum-scaling} and \eqref{eq:4D-dilaton-D1} we see that
\begin{equation}
    \left( \frac{M_{\textrm{KK}}}{\MPlfour} \right)^{2} = \frac{1}{\left( \mathcal{V}_{\gamma_{0}}^{\IIA} \right)^{\frac{2}{3}}} \left( \frac{M_{s}}{\MPlfour} \right)^{2} \sim \frac{1}{\lambda} \xrightarrow{\QC} \frac{1}{\lambda^{3}}\,.
\end{equation}
Along the D4-limit the only cycles whose volume varies are the 3-cycles. Since we only have even D-branes available, no other candidate tower except for that of D0-branes exists that could signal a decompactification to M-theory. Employing \eqref{eq:D4-gIIA} and \eqref{eq:4D-dilaton-D1} we observe that their mass scales like
\begin{equation}
    \left( \frac{M_{\textrm{D0}}}{\MPlfour} \right)^{2} = \frac{1}{g_{\IIA}^{2}} \left( \frac{M_{s}}{\MPlfour} \right)^{2} \sim 1 \xrightarrow{\QC} 1
\end{equation}
and hence we are safely within the realm of validity of type IIA string theory. The mirror tower of particles in the D1-limit is given by D3-branes wrapping the mirror dual to the SYZ-cycle which, although $\Omega_{Y}$ is constant along the D1-limit, has a varying volume due to the factor of $\mathcal{V}_{Y}^{\IIB}$ in the analogous expression to \eqref{eq:3-cycle-vol}.

\vspace{\baselineskip}
\textbf{A-F1-limit:} In the mirror dual to the F1-limit, which we will refer to as the A-F1-limit, no D-instantons contribute significantly and the hypermultiplet moduli space is $\alpha'$-exact. As discussed before, quantum corrections imply the freezing of the complex structure moduli. Taking this into account, as well as \eqref{eq:mirror-map} and \eqref{eq:SYZ-vol}, we obtain
\begin{equation}
    g'_{\IIA} \sim \frac{g'_{\IIB}}{(t^{'a})^{\frac{3}{2}}} \sim \frac{1}{\lambda^{\frac{3}{4}}} \xrightarrow{\QC} \frac{1}{\lambda^{\frac{3}{2}}}\,.
\label{eq:A-F1-gIIA}
\end{equation}
The lightest string is the fundamental type IIA string, whose tension behaves, in view of \eqref{eq:4D-dilaton-F1}, like
\begin{equation}
    \frac{T_{\textrm{F1}}}{\left( \MPlfour \right)^{2}} = \left( \frac{M'_{s}}{\MPlfour} \right)^{2} \sim \frac{1}{\lambda^{\frac{3}{2}}} \xrightarrow{\QC} \frac{1}{\lambda^{3}}\,.
\end{equation}
In Appendix \ref{app:A} we argue that the KK scale is set by the shrinking $S^{3}$. This scale tends to zero at the same rate as the fundamental string scale once the periods have approached their constant value
\begin{equation}
    \left( \frac{M_{\textrm{KK}}}{\MPlfour} \right)^{2} = \frac{1}{\left( \mathcal{V}_{\gamma_{0}}^{'\IIA} \right)^{\frac{2}{3}}} \left( \frac{M'_{s}}{\MPlfour} \right)^{2} \sim \frac{1}{\lambda} \xrightarrow{\QC} \frac{1}{\lambda^{3}}\,,
\end{equation}
where we have used \eqref{eq:SYZ-vol} and \eqref{eq:4D-dilaton-F1}. The mass scale of the D0-branes, obtained from \eqref{eq:A-F1-gIIA} and \eqref{eq:4D-dilaton-F1}, remains constant along the limit,
\begin{equation}
    \left( \frac{M_{\textrm{D0}}}{\MPlfour} \right)^{2} = \frac{1}{g_{\IIA}^{'2}} \left( \frac{M'_{s}}{\MPlfour} \right)^{2} \sim 1 \xrightarrow{\QC} 1\,,
\end{equation}
and therefore we are safely within the type IIA framework.

To conclude, on the type IIB side the self-similarity of the theory under strong coupling limits was instrumental in relating the D1- and F1-limits. Although ten-dimensional type IIA string theory does not enjoy such an S-duality, and just decompactifies to M-theory under purely strong coupling limits, when compactified on a Calabi-Yau threefold it inherits the S-duality of type IIB string theory through mirror symmetry. Therefore, the D4- and A-F1-limits are related to each other in this way, but the possibility to connect the two frames is a property of the compactified theory only.

\subsection{Identification of the moduli spaces}
\label{subsec_Id}

As mentioned at the beginning of Section \ref{sec:type-II-results}, the hypermultiplet moduli space of five-dimensional M-theory and that of four-dimensional type IIA string theory can be identified. This owes to the fact that when compactifying five-dimensional M-theory on $S^{1}$ the M-theory circle is associated with a vector multiplet from the four-dimensional M-theory perspective.
More precisely, the radius in five-dimensional Planck units is expressed in terms of the volume of $X$ in string units via the standard relation
\begin{equation}
    R \, \MPlfive = (\Vfour)^{\frac{1}{3}}\,.
\end{equation}

Therefore, the hypermultiplet moduli space of M-theory on $\mathbb{R}^{1,4} \times X$ remains unaltered if we further compactify to obtain M-theory on $\mathbb{R}^{1,3} \times X \times S^{1}$, which can then be identified with type IIA string theory on $\mathbb{R}^{1,3} \times X$.

The precise equivalence between the M-theory and type IIA quantities can be seen explicitly from the dimensional reduction of eleven-dimensional supergravity. By comparison of the Lagrangian densities in \cite{Cadavid:1995bk}, where the aforementioned reduction was explicitly carried out, one infers that the complex structure moduli $z^{a}_{\IIA}$ remain the same on both sides, while the volume scalar in M-theory gets identified with
\begin{equation}
    \Veleven = \frac{\Vfour}{g_{\IIA}^{2}}\,.
\end{equation}
This is essentially the four-dimensional dilaton
\begin{equation}
    \frac{\Vfour}{g_{\IIA}^{2}} = \frac{\Vfour}{g_{\IIB}^{2}} \frac{1}{\left( \mathcal{V}_{\SYZ}^{\IIA} \right)^{2}} \sim \frac{\int_{X} \Omega_{X} \wedge \overline{\Omega}_{X}}{g_{\IIB}^{2}} \sim \frac{\mathcal{V}_{Y}(t^{a})}{g_{\IIB}^{2}} \sim e^{-2\varphi_{4}}\,.
\end{equation}
Therefore, the coordinates employed in all three hypermultiplet moduli spaces considered are equivalent after taking into account circle reduction and mirror symmetry, so that we can directly translate the results on quantum corrections from one setting to another. The volumes of the three-cycles are related, via \eqref{eq:3-cycle-vol}, as
\begin{equation}
    \mathcal{V}_{\textrm{3-cycle}}^{\textrm{11D}} = \frac{\mathcal{V}_{\textrm{3-cycle}}^{\IIA}}{g_{\IIA}}\,.
\label{eq:3-cycle-M-to-A}
\end{equation}

We have seen that the D4- and A-F1-limits are truly four-dimensional physical settings that do not undergo a decompactification to M-theory, where a rescaling of the K\"ahler forms like $J_{\textrm{M}} = J^{\IIA}/g_{\IIA}^{2/3}$ would be justified as exploited in \cite{Lee:2019wij}. Similarly, the classical membrane limits considered in Section \ref{sec:cl-membrane-lims} are five-dimensional scenarios in which the M-theory circle is not present. Nonetheless, the identifications just discussed allow us to formally express all the mass scales of the classical membrane limits in terms of the type IIA variables and then, through mirror symmetry, in terms of the type IIB quantities discussed in \cite{Marchesano:2019ifh,Baume:2019sry}. For the latter the quantum corrections are known, a fact that we exploit in the next section.

\section{Quantum obstructions to membrane limits}
\label{sec:quantum-obstruction}

We are finally in a position to address the main question of this article, namely
the fate of the classical membrane limits in five-dimensional M-theory, as introduced in Section \ref{sec:cl-membrane-lims}.
Recall that in these limits a ``critical'' membrane, either the M2-brane or the M5-brane wrapping a shrinking $T^3$ fibre, classically becomes light at the same rate as the KK modes. We refer to these two classical limits as the M2- and M5-limit, respectively. With the help of the results from Section \ref{sec_IIAandM} we will now show that quantum corrections obstruct such membrane limits, precisely as predicted on general grounds in Section \ref{sec:dimensional_reduction}.

\subsection{M5-limit}

The classical M5-limit occurs when either of the two conditions in \eqref{eq:M5-condition} is fulfilled. The first of them, when expressed in type IIA variables using \eqref{eq:3-cycle-M-to-A}, reads
\begin{equation}
    \vol[\IIA]{\mathcal{A}} \sim \frac{g_{\IIA}^{2}}{\vol[\IIA]{\mathcal{B}}}\,,
\label{eq:M5-condition-1}
\end{equation}
while the second one is
\begin{equation}
    \vol[\IIA]{\mathcal{A}} \sim \frac{g_{\IIA}^{2}}{\left( \Vfour \right)^{\frac{1}{2}}}\,.
\label{eq:M5-condition-2}
\end{equation}

Let us focus on \eqref{eq:M5-condition-1} first. Consider a situation in which the mirror manifold $Y$ is homogeneously growing (in string units), i.e.\ where along the trajectory in hypermultiplet moduli space, expressed in terms of the type IIB variables, all K\"ahler coordinates scale like $\lambda$, with $\lambda \rightarrow \infty$. From the form of the periods of $X$ under mirror symmetry it follows that the 3-cycles whose volume exhibits the slowest and fastest scaling are given by the SYZ-cycle $\gamma^{0}$ with the property \eqref{eq:SYZ-vol} and its symplectic dual $\gamma_{0}$ scaling as in \eqref{eq:SYZ-pair-vol}, as we saw in the D4-limit. This means that the condition \eqref{eq:M5-condition-1} translates into 
\begin{equation}
    g_{\IIA} \sim 1\,,
\end{equation}
which is the behaviour classically found for \eqref{eq:D4-gIIA}. Therefore, under these conditions the M5-limit corresponds to the classical trajectory of the D4-limit, or its mirror dual, the D1-limit. After taking the quantum corrections into account, we find that the relevant scales now behave like
\begin{align}
    \frac{T_{\textrm{M2}}}{\left( \MPlfive \right)^{3}} &\sim \frac{g_{\IIA}^{2}}{\Vfour} \sim 1 \xrightarrow{\QC} \frac{1}{\lambda^{\frac{3}{2}}} \sim \frac{1}{\tilde{\lambda}}\,, \label{eq:M5-mass-scales-1}\\
    \frac{T_{\textrm{M5}|_{\mathcal{A}}}}{\left( \MPlfive \right)^{3}} &\sim \frac{\vol[\IIA]{\mathcal{A}} g_{\IIA}}{\Vfour} \sim \frac{1}{\lambda^{\frac{3}{2}}} \xrightarrow{\QC} \frac{1}{\lambda^{3}} \sim \frac{1}{\tilde{\lambda}^{2}}\,, \label{eq:M5-mass-scales-2}\\
    \left( \frac{M_{\textrm{KK,}\mathcal{B}}}{\MPlfive} \right)^{3} &\sim \frac{g_{\IIA}^{3}}{\vol[\IIA]{\mathcal{B}} \Vfour} \sim \frac{1}{\lambda^{\frac{3}{2}}} \xrightarrow{\QC} \frac{1}{\lambda^{\frac{9}{2}}} \sim \frac{1}{\tilde{\lambda}^{3}}\,, \label{eq:M5-mass-scales-3}\\
    \left( \frac{M_{\textrm{KK,}\mathcal{V}}}{\MPlfive} \right)^{3} &\sim \left( \frac{g_{\IIA}^{2}}{\Vfour} \right)^{\frac{3}{2}} \sim 1 \xrightarrow{\QC} \frac{1}{\lambda^{\frac{9}{4}}} \sim \frac{1}{\tilde{\lambda}^{\frac{3}{2}}}\,, \label{eq:M5-mass-scales-4}
\end{align}
with $\tilde{\lambda} \rightarrow \infty$. The impact of the quantum corrections was characterised in the type IIB language as the freezing of the D1-instanton contributions. By \eqref{eqD1inst} this translates, on the type IIA side, into the freezing of the D2-instanton contributions associated with (1,2)-cycles. In M-theory, finally, the quantum corrections hence freeze the volumes of said (1,2)-cycles measured in eleven-dimensional units. The classical membrane limit is deflected by the effect of quantum corrections coming from M2-brane instantons on the 3-cycles. The five-dimensional M-theory undergoes a decompactification limit along the trajectory and realises exactly the ratio \eqref{eq:membrane-KK-ratio} between the membrane tension and the KK scale which is predicted by requiring consistency under dimensional reduction. In Section \ref{sec_decomp} we will further analyse the decompactification limit induced by the quantum corrections

Let us briefly comment on the second putative realisation of an M5-limit, corresponding to \eqref{eq:M5-condition-2}.
This condition results, for the mirror of a homogeneously growing manifold, in a scaling
\begin{equation}
    g_{\IIA} \sim \frac{1}{\lambda^{\frac{3}{4}}}\,.
\label{eq:M5-condition-2-lambda}
\end{equation}
At the classical level the tension of the wrapped M5-brane indeed sits at the KK scale set by the overall volume of the manifold, but the assumption that this was the lowest-lying KK scale was unfounded:
\begin{align}
    \frac{T_{\textrm{M2}}}{\left( \MPlfive \right)^{3}} &\sim \frac{g_{\IIA}^{2}}{\Vfour} \sim \frac{1}{\lambda^{\frac{3}{2}}}\,,\\
    \frac{T_{\textrm{M5}|_{\mathcal{A}}}}{\left( \MPlfive \right)^{3}} &\sim \frac{\vol[\IIA]{\mathcal{A}} g_{\IIA}}{\Vfour} \sim \frac{1}{\lambda^{\frac{9}{4}}}\,,\\
    \left( \frac{M_{\textrm{KK,}\mathcal{B}}}{\MPlfive} \right)^{3} &\sim \frac{g_{\IIA}^{3}}{\vol[\IIA]{\mathcal{B}} \Vfour} \sim \frac{1}{\lambda^{\frac{15}{4}}}\,,\\
    \left( \frac{M_{\textrm{KK,}\mathcal{V}}}{\MPlfive} \right)^{3} &\sim \left( \frac{g_{\IIA}^{2}}{\Vfour} \right)^{\frac{3}{2}} \sim \frac{1}{\lambda^{\frac{9}{4}}}\,.
\end{align}
Already at the classical level do we therefore encounter a decompactification limit. 
This is, in fact, no different than in the analogous type IIB limit: there the condition \eqref{eq:M5-condition-2-lambda} would imply that
\begin{equation}
    g_{\IIB} = \frac{g_{\IIA}}{\mathcal{V}_{\SYZ}^{\IIA}} \sim \lambda^{\frac{3}{4}} \lesssim \lambda \sim t^{a}\,,
\end{equation}
and therefore we also face a classical decompactification limit from the point of view of the type II theories.

\subsection{M2-limit}

The classical M2-limit corresponds to the condition \eqref{eq:M2-condition}, i.e.\ the KK tower becoming light at the fastest rate must be associated with a 3-cycle that is not scaling in eleven-dimensional Planck units. Naively, one might think that a classical M2-limit cannot be realised since as soon as (measured in eleven-dimensional units) a 3-cycle shrinks the wrapped M5-brane will be lighter, while if it grows the corresponding KK tower will be leading.

However, in view of the M5-limit discussed above, it is natural to expect the M2-limit to correspond to the A-F1-limit. This is indeed the case, and the KK towers that would naively lead to a decompactification in the M2-limit are precisely the ones argued to be absent in Appendix \ref{app:A}. Taking therefore only the relevant scales into account\footnote{The arguments after (\ref{Mkkgammaalpha}) also explain why there are no KK states in the classical theory associated with the scale of $\Veleven$, which would naively destroy the membrane limit already before quantum corrections come into play.} and expressing the A-F1-limit in the M-theory language, we find the scaling behaviour
\begin{align}
    \frac{T_{\textrm{M2}}}{\left( \MPlfive \right)^{3}} &\sim \frac{g_{\IIA}^{2}}{\Vfour} \sim \frac{1}{\lambda^{\frac{3}{2}}} \xrightarrow{\QC} \frac{1}{\lambda^{3}} \sim \frac{1}{\tilde{\lambda}^{2}}\,, \label{eq:M2-mass-scales-1}\\
    \frac{T_{\textrm{M5}|_{\mathcal{B}}}}{\left( \MPlfive \right)^{3}} &\sim \frac{\vol[\IIA]{\mathcal{B}} g_{\IIA}}{\Vfour} \sim 1 \xrightarrow{\QC} \frac{1}{\lambda^{\frac{3}{2}}} \sim \frac{1}{\tilde{\lambda}}\,, \label{eq:M2-mass-scales-2}\\
    \left( \frac{M_{\textrm{KK,}\mathcal{A}}}{\MPlfive} \right)^{3} &\sim \frac{g_{\IIA}^{3}}{\vol[\IIA]{\mathcal{A}} \Vfour} \sim \frac{1}{\lambda^{\frac{3}{2}}} \xrightarrow{\QC} \frac{1}{\lambda^{\frac{9}{2}}} \sim \frac{1}{\tilde{\lambda}^{3}}\,, \label{eq:M2-mass-scales-3}\\
    \left( \frac{M_{\textrm{KK,}\mathcal{V}}}{\MPlfive} \right)^{3} &\sim \left( \frac{g_{\IIA}^{2}}{\Vfour} \right)^{\frac{3}{2}} \sim \frac{1}{\lambda^{\frac{9}{4}}} \xrightarrow{\QC} \frac{1}{\lambda^{\frac{9}{2}}} \sim \frac{1}{\tilde{\lambda}^{3}}\,, \label{eq:M2-mass-scales-4}
\end{align}
with $\tilde{\lambda} \rightarrow \infty$. Note that here the $\mathcal{A}$-cycle and $\mathcal{B}$-cycle are the $S^{3}$ and $T^{3}$ respectively, since the dependence of the volumes on $\lambda$ is the inverse of that for the D4-limit. Once again, the classical membrane limit is deflected by the quantum corrections and we observe the ratio \eqref{eq:membrane-KK-ratio} between the scalings, as expected from consistency under dimensional reduction.

We might wonder what is the fate of the membrane coming from wrapping an M5-brane on the shrinking $S^{3}$. After taking the quantum corrections into account it falls behind,
\begin{equation}
    \frac{T_{\textrm{M5}|_{\mathcal{A}}}}{\left( \MPlfive \right)^{3}} \sim \frac{\vol[\IIA]{\mathcal{A}} g_{\IIA}}{\Vfour} \sim \frac{1}{\lambda^{\frac{3}{2}}} \xrightarrow{\QC} \frac{1}{\lambda^{\frac{3}{2}}} \sim \frac{1}{\tilde{\lambda}}\,,
\end{equation}
but in the classical limit we might be tempted to analyse its role as it sits at the same scale as the fundamental string. We could have asked this already for the D4-brane wrapping the same cycle in the A-F1-limit. The resulting object corresponds to a non-critical string and therefore does not lead to a competing critical string, even in the classical theory. This non-critical string has an analogue also in the D1-limit: indeed, the D4-brane wrapping the $S^{3}$ in the A-F1-limit dualises to a D7-brane wrapping the whole Calabi-Yau $Y$ in the F1-limit, which in turn is S-dual to a $(0,1)$ 7-brane wrapping the whole Calabi-Yau $X$ in the D1-limit.

\subsection{Decompactification process}\label{sec_decomp}

We now analyse in more detail the process of decompactification characterising the quantum deflected membrane limits. For concreteness, the discussion is phrased in the framework of the M5-limit fulfilling condition \eqref{eq:M5-condition-1}.

With the quantum corrections taken into account, the KK tower from the $S^{3}$ base of the SYZ fibration becomes light at the fastest rate. The theory should therefore undergo a decompactification to eight-dimensional M-theory with the internal dimensions accounted for by the $T^{3}$. As the volume of the base grows without bound all supersymmetry breaking defects, in particular the degeneration loci of the $T^3$ fibration, are driven to infinity, thereby restoring the appropriate amount of supersymmetry.

In the eight-dimensional theory we measure the mass scales in terms of $\MPleight$, finding
\begin{align}
    \frac{T_{\textrm{M}2}}{\left( \MPleight \right)^{3}} &\sim \left( \frac{1}{\vol[\textrm{11D}]{\mathcal{A}}} \right)^{\frac{1}{2}} \sim \lambda^{\frac{3}{4}}\,,\\
    \frac{T_{\textrm{M}5|_{\mathcal{A}}}}{\left( \MPleight \right)^{3}} &\sim \left( \vol[\textrm{11D}]{\mathcal{A}} \right)^{\frac{1}{2}} \sim \frac{1}{\lambda^\frac{3}{4}}\,.
\end{align}
If we now consider the tower of particles coming from M2-branes wrapped on $S^{1} \times S^{1} \subset T^{3}$ we see that they signal further decompactification to eleven dimensions, becoming light like
\begin{equation}
    \frac{T_{\textrm{M}2|_{S^{1} \times S^{1}}}}{\MPleight} \sim \left( \vol[\textrm{11D}]{\mathcal{A}} \right)^{\frac{3}{2}} \sim \frac{1}{\lambda^{\frac{9}{4}}}\,.
\end{equation}
These set the KK scale
\begin{equation}
    \tilde{M}_{\textrm{KK}} := T_{\textrm{M}2|_{S^{1} \times S^{1}}}
\end{equation}
for a dual torus $\tilde{T}^{3}$ with volume
\begin{equation}
    \textrm{vol}(\tilde{T}^{3}) = \left( \frac{1}{\tilde{M}_{\textrm{KK}}} \right)^{3} \sim \left( \frac{1}{(\vol[\textrm{11D}]{\mathcal{A}})^\frac{2}{3}} \MPleleven \right)^{3}\,.
\end{equation}
Compactifying the dual eleven-dimensional theory on it, we obtain the relation
\begin{equation}
    \left( \tilde{M}_{\textrm{Pl}}^{\textrm{11D}} \right)^{3} = \left( \MPleight \right)^{2} \MPleleven \left( \vol[\textrm{11D}]{\mathcal{A}} \right)^{\frac{2}{3}},
\end{equation}
from which one reads
\begin{equation}
    \left( \frac{\MPleight}{\tilde{M}_{\textrm{Pl}}^{\textrm{11D}}} \right)^{3} \sim \left( \frac{1}{\vol[\textrm{11D}]{\mathcal{A}}} \right)^\frac{1}{2} \sim \lambda^\frac{3}{4}\,.
\end{equation}
Expressing the tensions in terms of $\tilde{M}_{\textrm{Pl}}^{\textrm{11D}}$ we find
\begin{equation}
    \frac{T_{\textrm{M}2}}{\left( \tilde{M}_{\textrm{Pl}}^{\textrm{11D}} \right)^{3}} \sim \lambda^{\frac{3}{2}}\,,\quad \frac{T_{\textrm{M}5|_{\mathcal{A}}}}{\left( \tilde{M}_{\textrm{Pl}}^{\textrm{11D}} \right)^{3}} \sim 1\,.
\end{equation}
We conclude that the wrapped M5-brane is the new M2-brane while the M2-brane orthogonal to the original torus becomes the M5-brane wrapping the dual torus. 

\section{Conclusions}
\label{sec:conclusions}

In five-dimensional M-theory we were able to engineer classical infinite distance limits in the hypermultiplet moduli space in which a critical membrane, in the terminology of Section \ref{sec:dimensional_reduction}, becomes parametrically light at the same rate as the KK scale. These trajectories are equivalent, under identification of the hypermultiplet moduli spaces of M-theory on $\mathbb{R}^{1,4} \times X$ and type IIA string theory on $\mathbb{R}^{1,3} \times X$ and application of mirror symmetry, to the classical string limits discussed in \cite{Marchesano:2019ifh,Baume:2019sry}. In these an emergent critical string becomes tensionless parametrically faster than the KK scale.

Taking quantum corrections into account modifies the string limits such that the tension of the critical string becoming parametrically light is bounded from below by the KK scale. Translating the corrections to the trajectories to the M-theory setting, we see that the membrane limits are deflected into a limit with scaling 
\begin{align}
    \frac{T_{\textrm{heavy}}}{\left( \MPlfive \right)^{3}} &\sim \frac{1}{\tilde{\lambda}}\,,\\
    \frac{T_{\textrm{light}}}{\left( \MPlfive \right)^{3}} &\sim \frac{1}{\tilde{\lambda}^{2}}\,,\\
    \left( \frac{M_{\textrm{KK}}}{\MPlfive} \right)^{3} &\sim \frac{1}{\tilde{\lambda}^{3}}\,,
\end{align}
with $\tilde{\lambda} \rightarrow \infty$. This explicitly reproduces the behaviour expected from consistency under dimensional reduction, as obtained in \eqref{eq:membrane-KK-ratio}.

The identification of the moduli spaces relied on the fact that, from the point of view of five-dimensional M-theory compactified on $S^{1}$, the M-theory circle is a vector multiplet. As a consequence, the hypermultiplet moduli space of M-theory on $\mathbb{R}^{1,4} \times X$ and M-theory on $\mathbb{R}^{1,3} \times X \times S^{1}$ are identical, the latter theory being equivalent to type IIA string theory on $\mathbb{R}^{1,3} \times X$.

Both in five-dimensional M-theory and in type IIA string theory we are considering pure hypermultiplet moduli space trajectories. The complex structure moduli are the same in both theories, while the remaining coordinate is the volume scalar for M-theory and the four-dimensional dilaton for string theory. This four-dimensional dilaton involves the string coupling and the volume of $X$ measured in string units, which from the type IIA perspective is unchanged as no K\"ahler moduli are varied. Therefore, the scaling of the volume scalar measured in eleven-dimensional units directly determines the scaling of the type IIA string coupling.

Type IIA hypermultiplet and vector multiplet moduli spaces locally factor, and the possible limits in the vector multiplet moduli space were studied in \cite{Lee:2019wij}, where no membrane limit was found. The light membranes in the five-dimensional theory would become, upon wrapping this M-theory circle, the light strings in the type IIA string theory, and their mass scale would appropriately pick up a factor of the M-theory circle radius. Therefore, we can interpret the fact that five-dimensional M-theory membrane limits turn into decompactification limits after quantum corrections are taken into account as a preventive measure against pathological string limits in the related, but not physically equivalent, four-dimensional string theory settings. In other words, the decoupling of membranes in the M-theory limits is necessary for the consistency under dimensional reduction of the Emergent String Conjecture.

Indeed, all compactifications of a consistent theory of quantum gravity must be consistent as well. Eleven-dimensional M-theory gives rise to both the five-dimensional M-theory and the four-dimensional type IIA string theory as considered in this article. The spectra of light states along the hypermultiplet moduli space limits of these two descendant theories are connected so as to ensure the quantum consistency of both. 

Without the hypermultiplet moduli space identification there is no obvious \textit{a priori} reason for the separation of scales \eqref{eq:membrane-KK-ratio} between the KK modes and the lightest (critical) membrane in the theory, as observed in five-dimensional M-theory.
The consistency of the Emergent String Conjecture under dimensional reduction provides a rationale for this relation and hence sheds new light also on the asymptotics of quantum gravity theories with no critical strings in their spectrum.

\section*{Acknowledgements}
This work was inspired by the discussions in the Swampland research seminar within the Quantum Universe Cluster of Excellence in Hamburg. We thank all participants, especially Murad Alim, Vicente Cort\'es, Nicole Righi, Arpan Saha, J\"org Teschner, Ivan Tulli and Alexander Westphal, for many insightful exchanges of ideas. We are also grateful to Ivan Tulli and Max Wiesner for comments on the draft. T.W. thanks Seung-Joo Lee and Wolfgang Lerche for important discussions and for collaboration on related topics. The authors are supported in part by Deutsche Forschungsgemeinschaft under Germany's Excellence Strategy EXC 2121 Quantum Universe 390833306.

\appendix

\section{\label{app:A}KK scale in A-F1-limits}

In this Appendix, we comment on the correct identification of the KK scale in the A-F1-limit of Section \ref{sec:mirror-map}.
As claimed there, the KK scale is set by the modes coming from the shrinking $S^{3}$, which is not the fastest growing cycle and in fact is shrinking along the limit. To understand this, we first emphasize that the truly meaningful scalings are the quantum-corrected ones, since in the A-F1-limit the distinction between classical and quantum-corrected trajectories is arbitrary.\footnote{See footnote \ref{fn:classical-quantum}.} The rate at which all possible KK modes become light is uniform after taking the corrected trajectories into account. Indeed, deep enough along the limit eventually $z^{a}_{\IIA} \sim \textrm{const.}$, and therefore we can directly write
\begin{equation}
    \left( \frac{M_{\textrm{KK}}}{\MPlfour} \right)^{2} = \frac{1}{\left( \mathcal{V}_{\textrm{3-cycle}}^{'\IIA} \right)^{\frac{2}{3}}} \left( \frac{M'_{s}}{\MPlfour} \right)^{2} \xrightarrow{\QC} 1 \cdot \frac{1}{\chi'} \sim \frac{1}{\lambda^{3}}\,,
\end{equation}
without reference to the specific 3-cycle considered.

However, at least for the classical A-F1-limit, we find some additional KK towers with scaling
\begin{equation}
\label{Mkkgammaalpha}
    \left( \frac{M_{\textrm{KK},\gamma^{\alpha}}}{\MPlfour} \right)^{2} = \frac{1}{\left( \mathcal{V}_{\gamma^{\alpha}}^{'\IIA} \right)^{\frac{2}{3}}} \left( \frac{M'_{s}}{\MPlfour} \right)^{2} \sim \frac{1}{\lambda^{2-\frac{\alpha}{3}}}\,,
\end{equation}
where $\alpha = 0,1,2,3$. The case with $\alpha = 3$ corresponds to the $S^{3}$ tower highlighted above, but, for example, the putative KK tower associated with the growing $T^{3}$ ($\alpha = 0$) becomes light faster than any other scale in the problem. This cannot be true, as no such phenomenon occurs in the mirror dual type IIB limit, i.e. in the classical F1-limit.

To resolve this puzzle, recall that we are using the SYZ fibration structure and the LCS behaviour of the periods in order to extract the scalings for the classical limits. This works well in the limit of large $S^{3}$ base, but soon becomes invalid for a trajectory like the one taken in the A-F1-limit. Luckily, mapping the quantum-corrected F1-limit to the mirror side is still possible thanks to the simplicity provided by the freezing of the K\"ahler coordinates on the type IIB side.

In spite of this, taking the classical limit precisely neglects these subtleties and exploits the LCS results beyond their regime of applicability. This is the origin of the pathological behaviour that is removed after taking the pertinent corrections into account. From this point of view, we might want to still argue why a tower like the one corresponding to the growing $T^{3}$ from the SYZ fibration should not be there even before exiting the regime of validity of the LCS approximation. A heuristic way to see this is to consider mirror symmetry as T-duality on the SYZ torus. This duality maps the $T^{3}$ KK tower observed on the type IIA side to string winding states along the dual SYZ torus fibre. If present, these light winding states would also endanger the classical F1-limit. However, on the IIB side it is clear that the winding states are not there as asymptotically light particles since the Calabi-Yau exhibits a non-trivial fibration structure rather than being a direct product $T^{3} \times S^{3}$. The degenerations of the $T^{3}$ fibre trivialise the homology class of the $S^{1}$ factors inside the $T^{3}$ as elements of $H_1(Y,\mathbb Z)$, giving a non-zero mass term to the naive winding states. As a consequence, the KK modes on the type IIA side must also be absent, or at least they cannot become asymptotically massless. This expected behaviour is confirmed by numerical computation of the eigenvalues of the scalar Laplace operator on the Dwork family $\left( \sum_i X_i^5 + \psi \prod_i X_i \right)$ of quintic threefolds~\cite{Braun:2008jp,Ashmore:2020ujw,Ashmore:2021qdf}. The towers corresponding to $\alpha = 1,2$ would correspond to cycles that are partially inside the $T^{3}$ fibre and degenerate in a similar fashion. Therefore, in the limits in which the manifold is shrinking from the type IIB perspective, we only take into account the $S^{3}$ KK modes for the classical analysis on the type IIA side.

\bibliography{references}
\bibliographystyle{utphys}

\end{document}